\documentclass[twocolumn,letterpaper,secnumarabic,amssymb, nobibnotes, aps, prl]{revtex4-1}        
\usepackage{epsfig}
\usepackage{dcolumn}
\usepackage{epstopdf}
\usepackage{graphicx}
\begin{document}

\title{Quantum body in uniform magnetic fields}  

 \author{S. Selenu}
\affiliation{}

\begin{abstract}
\noindent  In this article it will be presented the first attempt made  in order to perform gauge invariant calculations of  eigenstates of a quantum body in its condensed phase, the latter reacting to an external uniform magnetic field. The target is achieved introducing a new unitary translation operator transforming eigenstates  into a new set of eigenstates having different total linear momentum. This new quantum representation  solves the problem of calculating the magnetic response of quantum eigenstates of finite or either infinite periodic systems to uniform magnetic fields, where equivalence between the customarily used representation and the new representation has been made.
\end{abstract}

\date{\today} 
\maketitle

 \section{Introduction}

\noindent  Since last decades many attempts has been made in order calculating the response of matter compounds to external magnetic fields\cite{Vanderbilt,Resta} where also problems have been faced, in computational physics\cite{Martin},of calculating the orbital magnetisation of a quantum body\cite{Vanderbilt,Resta}. There has been the case of periodic boundary conditions applied to the system at null external fields in order reach a first clear understanding of the magnetism at the quantum level, but  still on debate, here differently it is put on relevance the use can be made of a  transformation of the eigenstates  via a new set of unitary operators translating the linear momentum of the set of eigenstates, the latter interacting with a magnetic field. The appearance of this unitary transformation allows  calculating directly the Hamiltonian of the system passing from a position dependent kinetic energy operator to an independent one\cite{Landau1,Zak,Resta} called boundary free,  showing the representability of the Hamiltonian being still in a gauge invariant form where measuring of the Hamiltonian operator can be brought forward in the linear momentum space. There  changing of its values can be directly measured by experimental means making the  theory here presented  extremely useful  when calculating eigenstates dependent on the wave vector ${\bf{k}}$. It is chosen eigenstates of matter being  in a Bloch form\cite{Bloch,Kittel,Ascroft,Resta,Vanderbilt} or either the new representation introduced  in\cite{selenu} making it worth calculating Hamiltonian eigenstates  for an  estimation of the quantum magnetization $\bf{M}$ of material compounds at the quantum level, wherethe latter  are dependent on external magnetic fields. In the first section will be dervived the unitary transformation of eigenstates of the quantal system, with the aim to introduce for the first time this  new representation and find in the second section a valuable  expression of an energy functional,  then allowing to calculate a quantum weighted magnetization. Conclusions are reported at the end of this article.\\\\\\
\section{Unitary Transformation of Quantum Eigenstates}
\label{hg}

In this section it will be derived an algorithm in order to model the quantum Hamiltonian of a quantum sample of matter being it in a  finite or infinite shape, still considering a  boundary free\cite{Resta,Vanderbilt} formalism when is the case of  the avoidance of the position operator by the kinetic energy part of the Hamiltonian. Differently by Zak models\cite{Zak}  here it is considered the set of quantum eigenstates of matter system being in a Bloch form not recurring to a representation where the position coordinates have a lack of a direct interpretation still considering anyhow eigenstates of matter being dependent on the wave vector ${\bf{k}}$.  It is worth considering ${\bf{k}}$ vectors of a quantum eigenstate of matter representing still the  linear momentum of the system\cite{Zak,Bloch,Landau1} where only the Bloch gauge is employed, i.e. where the form of the eigenstates is still represented as a phase factor multiplied by an eigen amplitude spanning the Hilbert space of the system:

\begin{eqnarray}
\label{equazionedel dipolo}
\Psi_{n,{\bf{k}}}=e^{i{\bf{k}}\cdot \bf{r}}u_{n,{\bf{k}}}
\end{eqnarray}

The unclear idea of a quasi momentum ${\bf{k}}$ required by Bloch and Zak\cite{Bloch,Zak} during their  first study of the transport of electrons in crystals, where  only periodicity is asked, makes difficult recognizing the behaviour of electrons should have on a finite sample where still eigen energies are amenable having dispersion relations dependent by the wave vector ${\bf{k}}$ of the state. Either in an amorphous system appears the ${\bf{k}}$  dependence of eigenstates as so  that the latter should not be confined to a prescribed unitary cell of a reciprocal space\cite{Ascroft} considering the sampling of the ${\bf{k}}$ vectors always due to a scattering process may affect the linear momentum of the system itself. Here considering the special case of a uniform magnetic field acting on a quantum body the wave vector should be scattered by a linear momentum, as it is well known\cite{Landau1,Jackson}, proportional to the potential vector ${\bf{A}}={\frac{e}{c}}\bf{B} \times \bf{r}$ of the system generating then on it a total linear momentum $\bar{P}$:
\noindent 
\begin{eqnarray}
\label{equazionedel}
\bar{P}=\langle \Psi{|\bf{p}} - {\frac{e}{c }}{\bf{A}}|\Psi \rangle
\end{eqnarray}

it making a position dependent  kinetic energy term in the Hamiltonian. The need of having a consistent unitary theory in order describe amorphous systems as well as periodic systems such as crystals makes then mandatory need of having an Hamiltonian operator represented by a boundary free  operator, i.e. not dependent on the shape of the matter compound under study. The target is reached  by use of a particular Unitary transformation $U$ of eigenstates $\Psi_{n,{\bf{k}}}$ that brings eigenstates at wave vector ${\bf{k}}$ to eigenstates of scattered linear momentum ${{\bf{k}}}-\frac{e}{c}\bf{A}$. The operator $U$ reads:

\noindent 
\begin{eqnarray}
\label{U}
U=e^{-i{\frac{e}{c}(\bf{B} \times \bf{i\nabla_{{\bf{k}}}})\cdot \bf{r}}}
\end{eqnarray}

Latter operator can be recasted as follows:

\begin{eqnarray}
\label{U1}
U=e^{i{\frac{e}{c}(\bf{B} \times \bf{r})\cdot {i\nabla_{{\bf{k}}}}}}
\end{eqnarray}

and by definition of a translation operator\cite{Landau1} it acts shifting the eigenstates from ${\bf{k}}$ to ${{\bf{k}}}-\frac{e}{c}\bf{B \times r}$ :

\begin{eqnarray}
\label{U2}
U\Psi_{n,{\bf{k}}}=e^{i{\frac{e}{c}(\bf{B} \times \bf{r})\cdot {i\nabla_{{\bf{k}}}}}}\Psi_{n,{\bf{k}}}=\Psi_{n,{\bf{k}}-\frac{e}{c}\bf{B \times r}}
\end{eqnarray}

making then possible evaluating at the knowledge of eigenstates $\Psi_{n,{\bf{k}}}$ the whole set of eigenstates $\Psi_{n,{\bf{k}}-\frac{e}{c}\bf{B \times r}} $ of the quantum body interacting with a uniform magnetic field $\bf{B}$. The Hamiltonian operator $H$ acting on $\Psi_{n,{\bf{k}}-\frac{e}{c}\bf{B \times r}}$  becomes:

\noindent 
\begin{eqnarray}
\label{equazione H}
H\Psi_{n,{\bf{k}}-\frac{e}{c}\bf{B \times r}}={[\frac{\bf{p}^2}{2m}+V(\bf{r})]}\Psi_{n,{\bf{k}}-\frac{e}{c}\bf{B \times r}}
\end{eqnarray}

where operator $H$ is formally transformed to  the Hamiltonian operator $\tilde{H}=U^+HU$,the latter acting on eigenstates $\Psi_{n,{\bf{k}}}$,  giving back the following eigenvalue equation: 

\noindent 
\begin{eqnarray}
\label{equazione H1}
\tilde{H}\Psi_{n,{\bf{k}}}={[\frac{({\bf{p}}-\frac{e}{c}\bf{B}\times i\nabla_{{\bf{k}}})^2}{2m}+V(\bf{r})]}\Psi_{n,{\bf{k}}}
\end{eqnarray}

In order to show the equivalence between eq.(\ref{equazione H1}) and the general Hamiltonian eigenvalue equation customarily employed in condensed matter physics, with the aim of  calculating eigenstates of matter interacting with a uniform magnetic field:
\noindent 
\begin{eqnarray}
\label{equazione Hp}
H\Psi_{n,{\bf{k}}}={[\frac{({\bf{p}}+\frac{e}{c}\bf{B}\times \bf{r})^2}{2m}+V(\bf{r})]}\Psi_{n,{\bf{k}}}
\end{eqnarray}

   it  is worth to  notice,  being  the shift of ${\bf{k}}$ to ${\bf{k}}-\frac{e}{c}\bf{B \times r}$, the eigenvalue equation for the quantum chiral eigenstate $\Psi$ is reached as:
\noindent 
\begin{eqnarray}
\label{equazione Hp1}
H\Psi_{n,{\bf{k}}-\frac{e}{c}\bf{B \times r}}={[\frac{({\bf{p}}+\frac{e}{c}\bf{B}\times \bf{r})^2}{2m}+V(\bf{r})]}\Psi_{n,{\bf{k}}}
\end{eqnarray}

by a direct shift of the reference system of wave vector ${\bf{k}}$ to ${\bf{k}}-\frac{e}{c}\bf{B \times r}$, then showing that a macroscopic energy observable is still definable and still invariant by the shifting of the quantum linear momentum in $\bf{k}$ space.  The latter enable us reaching a quantification of the amount of the gauge invariant energy of a magnetic systems,  solving then the targeted problem of calculating matter response to external uniform magnetic fields. In the next section it will be reported a derivation of the quantum energy functional should be employed in order calculate eigenstates of matter in a boundary free form while also deriving the expression of an Orbital Magnetisation. Conclusions are left to the last part of the article.

\section{Energy functional of a body in uniform magnetic fields}
In this section it will be introduced an energy functional, we shall perform a functional derivative,  in order  to obtain a set of quantum eigenvalue differential equations, that can be directly implemented in first principle  codes\cite{Martin} with the aim of modeling the ab initio  response of quantum matter to uniform magnetic fields. It is clear by the definition of the Hamiltonian operator reported in eq.(\ref{equazione H1}) that the energy observable   can be written in terms of powers of $\frac{1}{c}$ as in the  following:

\begin{eqnarray}
\label{equazione H01}
&&\langle \Psi_{n,{\bf{k}}}|\tilde{H}|\Psi_{n,{\bf{k}}}\rangle ={\langle \Psi_{n,{\bf{k}}}|[\frac{(\bf{p}-\frac{e}{c}\bf{B}\times i\nabla_{{\bf{k}}})^2}{2m}+V(\bf{r})]}|\Psi_{n,{\bf{k}}}\rangle  \\\nonumber
\end{eqnarray}
and the Hamiltonian operator becomes,
\begin{eqnarray}
\label{equazione H010}
&&{\tilde{H}=[\frac{{\bf{p}^2}}{2m}-\frac{e{\bf{p}}}{mc}\cdot {(\bf{B}\times i\nabla_{{\bf{k}} })}+ \frac{e^2}{2mc^2}{(\bf{B}\times i\nabla_{{\bf{k}}})^2}+V(\bf{r})]}
\end{eqnarray}

allowing to  evaluate the energy functional derivatives with respect to eigenstates of the body,  where are considered  variations of the total energy with respect to the quantum eigenfunctions $\Psi^*_{n,{\bf{k}}}$:

\begin{eqnarray}
\label{equazione H001}
\frac{\delta[ \langle \Psi | \tilde{H} |\Psi \rangle  -E\langle \Psi | \Psi \rangle]}{\delta \Psi^*}=0 \\\nonumber 
\end{eqnarray}

A linear  energy term can be evaluated,t due to the  magnetic field coupling with a quantum vector field $\bf{M}$, the latter  amenable of being recognized the quantum orbital magnetisation of matter\cite{Vanderbilt,Resta} where we expand the Hamiltonian operator only to first powers of $\frac{1}{c}$ reaching the following eigenvalue equation:

\begin{eqnarray}
\label{equazione H02}
\tilde{H}|\Psi_{n,{\bf{k}}}\rangle={[\frac{\bf{p}^2}{2m}-\frac{e{\bf{p}}}{mc}\cdot {(\bf{B}\times i\nabla_{{\bf{k}} })}+V(\bf{r})]}|\Psi_{n,{\bf{k}}}\rangle
\end{eqnarray}

Demonstration brought forward until now determines an expectation energy functional linear, in terms of the uniform magnetic field, then write the energy as a scalar product of the magnetic field and of a new field, we may recognize,  related to the orbital magnetization $\bf{M}$ of the body found in the case of null external uniform magnetic field\cite{Vanderbilt,Resta} where it is  the case of a crystal. In fact energy functional can be written as:

\begin{eqnarray}
\label{equazione H002}
{\langle \Psi_{n,{\bf{k}}}|\tilde{H}|\Psi_{n,{\bf{k}}}\rangle=\langle \Psi_{n,{\bf{k}}}|\frac{{\bf{p}^2}}{2m}|\Psi_{n,{\bf{k}}}\rangle+ E_{M,n} +\langle \Psi_{n,{\bf{k}}}|V(\bf{r})}|\Psi_{n,{\bf{k}}}\rangle\\\nonumber
\end{eqnarray}

where the amount of energy due to the magnetization of the body is then:
\begin{eqnarray}
\label{equazione H03}
&&E_{M,n}= \bf{B}\cdot \bf{M_n}\\\nonumber 
&&{\bf{M_n}}= \frac{e}{mc}\langle \Psi_{n,{\bf{k}}}|{\bf{p}\times i\nabla_{{\bf{k}} }}|\Psi_{n,{\bf{k}}}\rangle 
\end{eqnarray}

The evaluation of the energy term $E_{M,n}$ allows also having at hands a linear response of matter to magnetic fields and still define an orbital magnetization density either  in the general case of finite amorphous systems as an invariant trace of the vector  $M_n$ or either its weigthed average value as further  reported :
\begin{eqnarray}
\label{equazione H03}
&&{\bf{M}_{T}}=\frac{1}{V}\sum_n{\bf{M_n}}\\\nonumber
&&{\bf{M}}=\frac{1}{V}\sum_n f_n{\bf{M_n}}
\end{eqnarray}
being $f_n$ the occupation numbers of the eigenstate $n$ of the quantum body and $V$ its volume. The formula stays invariant either in the thermodynamic limit for infinite crystals or for the more appealing finite periodic systems routinely encountered in computational physics, then allowing to have a unifided theory making possible calculate magnetic properties of quantum matter or either compute them at the  first principles level in an abinitio framework\cite{Martin}. Put on forward these main results it will be reported conclusionsof the article in the next section.

\section{Conclusions}
In this article it has been introduced a new quantum representation  by the use of a gauge transformation of quantum eigenstates of matter in order to solve the problem of calculating the magnetic response of quantum eigenstates of finite or either infinite periodic systems to uniform magnetic fields. Equivalence between the customarily used representation and the new representation has been made, it  solving the problem of calculating quantum Hamiltonian operators in a boundary free context, i.e. having a form of the kinetic energy operator independent of the position operator,  it leaving to the freedom of calculating the magnetic response of matter in crystals as well as finite amorphous samples. This algorithm  may appear then useful in a first principles context where an ab initio modelling  of matter is made in order to simulate numerically the magnetic response of a quantum body to uniform magnetic fields\cite{Martin}.

\end{document}